\def \Rm {\mathbb R}

\def \Nm {\mathbb N}

\newcommand{\aver}[1]{\langle {#1} \rangle}
\newcommand{\Baver}[1]{\Big\langle {#1} \Big\rangle}

\newcommand{\bzero}{\mathbf 0}

   \newcommand{\bn}{\mathbf n}

\newcommand{\bx}{\mathbf x}

 \newcommand{\bF}{\mathbf F}

\newcommand{\mO}{\mathcal O}

\newcommand{\dif}{{\rm d}}

\newcommand{\rM}{{\rm M}}

\newcommand{\ra}{{\rm a}}

\newcommand{\rs}{{\rm s}}
\newcommand{\rt}{{\rm t}}

\newcommand{\bnabla}{\boldsymbol \nabla}
\newcommand{\bOmega}{\boldsymbol \Omega}

\newcommand{\Oxy}{{{\mathrm O}_2}}

\documentclass[preprint,12pt]{elsarticle}

\usepackage{amssymb}
\usepackage{amsthm}

\usepackage{lineno}

\usepackage{url,hyperref,microtype,subcaption}
\usepackage[onehalfspacing]{setspace}

\usepackage{amsmath,amsfonts,amssymb}
\usepackage{graphics,graphicx,color}
\usepackage{nicefrac}
\usepackage{natbib}

\journal{JQSRT}

\begin{document}

\begin{frontmatter}



\title{Moments of Sunlight Pathlength in Water and Aerosol Clouds from O$_2$ Spectroscopy: Exploitable Parameter Sensitivities}


\author[inst1]{Anthony B. Davis}

\affiliation[inst1]{
            organization={NASA Jet Propulsion Laboratory, California Institute of Technology},
            city={Pasadena},
            state={CA},
            country={United States} }

\author[inst2]{Quentin Libois}

\affiliation[inst2]{
            organization={Centre National de Recherche en Meteorologie CNRS, Universite de Toulouse, Meteo-France}, 
            city={Toulouse},
            country={France} }

\author[inst3]{Nicolas Ferlay}

\affiliation[inst3]{
            organization={Laboratoire d'Optique Atmospherique CNRS, Universite Lille-1}, 
            city={Villeneuve d'Ascq},
            country={France} }

\author[inst4]{Alexander Marshak}

\affiliation[inst4]{
            organization={NASA Goddard Space Flight Center, Climate and Radiation Laboratory},
            city={Greenbelt},
            state={MD},
            country={United States} }

\begin{abstract}
{\bf SYNOPSIS:}
According to the latest IPCC assessment, clouds and aerosols remain a major challenge in future climate prediction using Global Climate Models (GCMs). 
Consequently, NASA's 2017 Decadal Survey has made them, along with convection and precipitation, a high priority by calling them out as Designated Observables for future coordinated missions under the ``ACCP'' banner, now known as the {\it Atmospheric Observing System}.
Atmospheric science is now more than ever driving a renewal in remote sensing techniques that can probe clouds and aerosols more accurately and with improved sampling.
Herein, we bring new theoretical results that support a special form of Differential Optical Absorption Spectroscopy (DOAS) that uses oxygen absorption features in the visible/near-IR (VNIR) spectrum where there is an abundance of sunlight.
Having known concentration and cross-section, the remaining unknown in $\Oxy$ DOAS (DO$_2$AS) is the path that the light has followed through the absorbing gas.
In the presence of scattering by cloud and aerosol particles, that path is broken at each interaction.
Cumulative pathlength through the well-mixed $\Oxy$ gas thus becomes a continuous random variable, and its probability distribution function (PDF) will contain desirable information about the clouds and aerosols such as top and bottom altitudes of the layer they occupy as well as its optical thickness.
In this study, we build on previous work to compute statistical moments of the pathlength PDF.
Specifically, we show that mean and variance of the pathlength inside the scattering particulate medium convey different pieces of information, namely, size and opacity of the optical medium, hence geometric and optical thicknesses in the case of plane-parallel layers.
We also extend, with dense aerosol plumes in mind (e.g., wildfire smoke, volcanic ash, elevated dust), the close connection between geometric thickness and mean pathlength to the case of absorption by the scattering particles themselves.
In summary, we view pathlength moments as intermediate DO$_2$AS products that can be obtained from $\Oxy$ spectroscopic data, and we describe a preliminary algorithm to do just that.
In turn, these moments yield cloud or aerosol profile parameters of high interest: layer geometric and optical thicknesses.
One normally thinks about active sensors, radars and lidars, to do such atmospheric profiling.
In this case, the profile is parameterized and representative of some kind of horizontal average determined by the multiple scattering.
However, there is a clear advantage in using passive over active instrumentation, starting with the possibility of imaging over a large swath.
\end{abstract}


\begin{highlights}
\item 
New derivation of the ``4V/S'' invariance property of mean pathlength in arbitrarily-shaped optical media from first principles of radiative transfer.
\item 
Verification that, while mean pathlength conveys direct information about the size of the optical medium, pathlength variance increases with its opacity and, unlike reflectivity, without bound.
\item 
First prediction of how absorption by the scattering particles affects mean pathlength in plane-parallel media.
\item 
Preliminary algorithm for extracting pathlength moments from spectroscopic data.
\end{highlights}


\end{frontmatter}



\section*{Outline}
\label{sec:Intro}


In \S\ref{sec:InCloud_Paths}, we set up a theoretical and observational framework for relating DO$_2$AS at any spectral resolution to cloud properties through moments of the in-cloud pathlength distribution that is grounded in time-dependent 3D radiative transfer (RT). 
In \S\ref{sec:new_mean_derivation}, we present a new derivation for the mean pathlength in arbitrary cloud geometry.
In \S\ref{sec:diff_ppg}, new results for pathlength mean and variance in plane-parallel geometry are obtained.
On the one hand, we confirm that the mean is all about cloud size, variance informs us directly about cloud opacity in a regime where operational determinations of cloud optical thickness (COT) loose sensitivity.
An alternative approach for connecting 1st- and 2nd-order pathlength moments to cloud properties based on discrete-time random walk theory is presented in \ref{sec:Appendix_A}, thus reinforcing our findings.
On the other hand, with aerosol clouds in mind, we wonder how absorption by the scattering particles affects pathlength moments and quantify its impact on the mean.
In short, pathlength moments yield cloud properties, but it is not obvious how to go from DO$_2$AS observations to pathlength moments, so we describe a tentative algorithm in \ref{sec:Appendix_B}.

\section{Time-Dependent RT and Pathlength Moments}
\label{sec:InCloud_Paths}

Let $I(t,\bx,\bOmega)$ be the time-dependent specific intensity for a light beam intercepted at point $\bx$ propagating in direction $\bOmega$ expressed in [W/m$^2$/sr].
Rather than time $t > 0$ per se, we can use pathlength $ct$ cumulated since \emph{everywhere simultaneous} entrance into the arbitrarily-shaped optical medium M $\subset\Rm^3$ at its boundary $\partial$M, with $c$ denoting the speed of light in the optical medium.
Note that the physical units of $I(ct,\bx,\bOmega) = I(t,\bx,\bOmega)/c$ are [J/m$^3$/sr].
$I(ct,\bx,\bOmega)$ satisfies the integro-differential 3+1D RT equation (RTE):
\begin{equation} 
\label{eqn:3+1D_RTE}
   \left[ \frac{\partial\;}{\partial ct} + \bOmega\cdot\bnabla + \sigma(\bx) \right] \, I = \sigma_\rs(\bx) \int_{4\pi} P(\bOmega\cdot\bOmega^\prime) I(ct,\bx,\bOmega^\prime) \dif\bOmega^\prime,
\end{equation}
where $\sigma(\bx)$ is the extinction coefficient in [1/m], while $\sigma_\rs(\bx)$ is the scattering coefficient, with $P(\bOmega\cdot\bOmega^\prime)$ being the scattering phase function (PF) in [1/sr].
The PF is normalized as $2\pi\int_{-1}^{+1}P(\mu_\rs)\dif\mu_\rs = 1$, where $\mu_\rs = \bOmega\cdot\bOmega^\prime$ is the cosine of the scattering angle.
Sinks of radiant energy at point $(ct,\bx,\bOmega)$ in transport space are on the left-hand side of (\ref{eqn:3+1D_RTE}), namely advection and extinction, while sources are on the right-hand side, but only in-scattering of ambient light is considered in the present study.
Finally, a statement of initial ($t = 0$) and boundary ($\bx\in\partial\rM$) conditions for incoming radiation is required to define $I(t,\bx,\bOmega)$ uniquely everywhere and at all times.

We set the initial and boundary conditions (BCs) for the time-dependent 3D RTE (\ref{eqn:3+1D_RTE}) to
\begin{equation} 
\label{eqn:I_&_BCs}
  I(ct,\bx,\bOmega) = \delta(ct) E_0 / \pi, \text{ for } \bx\in\partial\rM \text{ and } \bOmega\cdot\bn(\bx) < 0,
\end{equation}
where $\delta(\cdot)$ is Dirac's delta function, $E_0$ is a uniform amount of radiant energy released into M per unit of surface on $\partial$M in [J/m$^2$], and division by $\pi$ accounts for the isotropic distribution of the incoming light.
Equation~(\ref{eqn:I_&_BCs}) describes a sudden, spatially and angularly uniform injection of light into the medium M.

Because of the multiplicity of possible paths, starting at the boundary, changing direction at every scattering, ending at $(\bx,\bOmega)$, after covering a distance $ct$, this pathlength can be viewed as a random variable. 
Its probability distribution function (PDF) in [1/m] is defined as 
\begin{equation} 
\label{eqn:ct_PDF}
  f(ct,\bx,\bOmega) = I(ct,\bx,\bOmega) \left/ \int_0^\infty I(ct,\bx,\bOmega) \dif ct \right.,
\end{equation}
which is dependent on position in transport space $(\bx,\bOmega)$.
From there, we can compute the $q^\text{th}$-order statistical moment of the random variable $ct$:
\begin{equation} 
\label{eqn:ct_moms_PDF}
  \Baver{(ct)^q}(\bx,\bOmega) = \int_0^\infty (ct)^q f(ct,\bx,\bOmega) \, \dif ct,
\end{equation}
where $q\in\Nm$ might need to have an upper bound to ensure convergence.
We are particularly interested in $q$ = 1,2.

Consider the Laplace transform of $f(ct,\bx,\bOmega)$:
\begin{equation} 
\label{eqn:ct_CF}
  \tilde{f}(s,\bx,\bOmega) = \int_0^\infty \exp(-s\, ct) \, f(ct,\bx,\bOmega) \, \dif ct = \aver{\exp(-s\, ct)}(\bx,\bOmega),
\end{equation}
which is non-dimensional. 
This is known as the characteristic (or ``moment-generating'') function of a PDF supported by $\Rm^+$, and it can be used to compute statistical moments using its successive derivatives at $s=0$:
\begin{align} 
\label{eqn:ct_moms_CF}
  \Baver{(ct)^q}(\bx,\bOmega) = {}& \left( -\frac{\partial \;}{\partial s} \right)^q \left. \tilde{f}(s,\bx,\bOmega) \right|_{s=0} \nonumber \\
                              = {}& \left( -\frac{\partial \;}{\partial s} \right)^q \left. \left. \tilde{I}(s,\bx,\bOmega) \right|_{s=0}
                             \right/ \tilde{I}(0,\bx,\bOmega).
\end{align}
These are \emph{non}-centered moments. 
There is also the ``cumulant-generating'' function $\log\tilde{f}(s,\bx,\bOmega)$ that is used to compute cumulants, i.e., the centered moments that are additive for sums of random variables.
Of particular interest here are the 1$^\text{st}$- and 2$^\text{nd}$-order cumulants, namely, the mean and variance of pathlength $ct$, respectively from $q = 1,2$:
\begin{align} 
\label{eqn:Avr_ct}
  \aver{ct}(\bx,\bOmega) = {}& \left. \left( -\frac{\partial \;}{\partial s} \right) \log\tilde{f}(s,\bx,\bOmega) \right|_{s=0} ; \\
\label{eqn:Var_ct}
  \text{Var}[ct](\bx,\bOmega) = \aver{(ct)^2}(\bx,\bOmega)-\aver{ct}^2(\bx,\bOmega) = {}& \left. \left( -\frac{\partial \;}{\partial s} \right)^2 \log\tilde{f}(s,\bx,\bOmega) \right|_{s=0}.
\end{align}

By Laplace transforming the 3+1D RTE in (\ref{eqn:3+1D_RTE}) term-by-term, we see that 
\begin{equation} 
\label{eqn:LT_I}
  \tilde{I}(s,\bx,\bOmega) \propto \int_0^\infty \exp(-s\, ct) \, I(ct,\bx,\bOmega) \, \dif ct
\end{equation}
obeys a steady-state 3D RTE:
\begin{equation} 
\label{eqn:3D_RTE}
   \left[ \bOmega\cdot\bnabla + (\sigma(\bx) + s) \right] \, \tilde{I} = \sigma_\rs(\bx) \int_{4\pi} P(\bOmega\cdot\bOmega^\prime) \tilde{I}(s,\bx,\bOmega^\prime) \dif\bOmega^\prime,
\end{equation}
irrespective of the proportionality constant that may be used in (\ref{eqn:LT_I}).
Note that the right-hand side of (\ref{eqn:LT_I}) is in [J/m$^2$/sr].
For instance, at $s = 0$, it expresses what portion of the original radiant energy release $E_0$ [J/m$^2$] in (\ref{eqn:I_&_BCs}) ends up at $(\bx,\bOmega)$, irrespective of the path $ct$ covered to get there.
At the same time, the steady-state radiance on the left-hand side of (\ref{eqn:LT_I}) is conventionally expressed in [W/m$^2$/sr] because of the steady sources.
This minor conundrum is resolved when we set the BCs for the steady-state problem further on.

In the above time-dependent 3D RTE (\ref{eqn:3+1D_RTE}), we can now identify $ct$ with the total pathlength $L$, and the Laplace conjugate variable $s$ in (\ref{eqn:3D_RTE}) with the spectral absorption coefficient $k_\lambda$ in [1/m] of an absorbing gas that uniformly fills the scattering medium M.
In that form, (\ref{eqn:LT_I}) becomes a predictive model for the spectral radiance $I_\lambda(\bx,\bOmega) \equiv \tilde{I}(k_\lambda,\bx,\bOmega)$ based on the radiance field $I(L,\bx,\bOmega)$ viewed as steady-state, but decomposed according to $L$, hence with units of [(W/m$^2$/sr)/m].
In that guise, Eq. (\ref{eqn:LT_I}) has been described as the ``equivalence'' theorem \citep{Irvine1964,Ivanov1972}.
We finally note that in (\ref{eqn:3D_RTE}), now viewed as a spectral RT problem, with $s \mapsto k_\lambda$, that quantity is no longer an independent variable.
Rather, $k_\lambda$ is a parameter---a given property of the optical medium.

\section{Universality of $\aver{ct}$ for Arbitrary Conservative Optical Media}
\label{sec:new_mean_derivation}

Returning to the time-dependent 3+1D RTE in (\ref{eqn:3+1D_RTE}), we integrate over $\bOmega$ in all $4\pi$ directions to obtain the continuity equation
\begin{equation} 
\label{eqn:continuity}
  \frac{\partial J}{\partial ct} + \bnabla\cdot\bF = -\sigma_\ra(\bx) \, J,
\end{equation}
where $\sigma_\ra(\bx) = \sigma(\bx)-\sigma_\rs(\bx)$ is the (particulate) absorption coefficient, or
\begin{equation} 
\label{eqn:conservation}
  \bnabla\cdot\tilde{\bF} = -\left( \sigma_\ra(\bx) + s \right)\,\tilde{J}
\end{equation}
in Laplace space, from (\ref{eqn:3D_RTE}), where
\begin{equation} 
\label{eqn:J_and_F_fluxes}
\begin{array}{ccc}
J(ct,\bx) = \int_{4\pi} I(ct,\bx,\bOmega)\,\dif\bOmega, & \bF(ct,\bx) = \int_{4\pi} \bOmega \, I(ct,\bx,\bOmega)\,\dif\bOmega, & \text{in }[\text{J/m}^3] \\
\tilde{J}(s,\bx) = \int_{4\pi} \tilde{I}(s,\bx,\bOmega)\,\dif\bOmega, & \tilde{\bF}(s,\bx) = \int_{4\pi} \bOmega \, \tilde{I}(s,\bx,\bOmega)\,\dif\bOmega, & \text{in }[\text{W/m}^2] \\
\end{array}
\end{equation}
in (\ref{eqn:continuity}) and (\ref{eqn:conservation}), respectively.
$J(ct,\bx)$ is the instantaneous \emph{scalar} (or actinic) flux, closely related to radiant energy \emph{density} $u(t,\bx) = J(ct,\bx)/c$, and $\bF(ct,\bx)$ is its \emph{vector} flux counterpart, a.k.a. radiant energy \emph{current}.
Their Laplace transforms, $\tilde{J}(s,\bx)$ and $\tilde{\bF}(s,\bx)$, behave like steady-state versions with variable amounts of gaseous absorption.
Equations (\ref{eqn:continuity}) and (\ref{eqn:conservation}) express the conservation of radiant energy, respectively, in the time- and Laplace domains.

We now apply the divergence theorem to the differentiable vector field $\tilde{\bF}(s,\bx)$:
\begin{equation} 
\label{eqn:divergence}
  \int_\rM \bnabla\cdot\tilde{\bF}\,\dif V(\bx) = \int_{\partial\rM} \tilde{\bF}\cdot\bn(\bx)\,\dif S(\bx),
\end{equation}
where $\bn(\bx)$ is the unitary outgoing normal vector at point $\bx$ on $\partial$M.
As is standard practice in radiative heating rate estimation, the integrand on the left-hand side can be replaced by $-\left( \sigma_\ra(\bx) + s \right)\tilde{J}(\bx)$, per radiant energy conservation expressed in (\ref{eqn:conservation}).

The Laplace transform of a delta function at the origin is unity.
Therefore, (\ref{eqn:I_&_BCs}) transforms into the (steady-state) BC
\begin{equation} 
\label{eqn:LT_BCs}
  \tilde{I}(s,\bx,\bOmega) = F_0 / \pi, \text{ for } \bx\in\partial\rM \text{ and } \bOmega\cdot\bn(\bx) < 0,
\end{equation}
where $F_0$ is a steady and uniform irradiance over $\partial$M expressed in the usual units of [W/m$^2$]. 
We thus ensure that $\tilde{I}(s,\bx,\bOmega)$ has the usual units of [W/m$^2$/sr].
In other words, the so-far unspecified proportionality constant in (\ref{eqn:LT_I}) is set to $F_0/E_0$ in [1/s].

The right-hand side of (\ref{eqn:divergence}) can be used to compute the mean in-cloud pathlength, spatially and angularly-averaged over $\partial$M and the 2$\pi$ solid angle for radiance contributing to the outgoing flux. Formally, 
\begin{equation} 
\label{eqn:avr_mean_ct}
  \aver{ct} = \left. \int_{\partial\rM} \int_{\bOmega\cdot\bn(\bx)>0} \aver{ct}(\bx,\bOmega) \times \bOmega\cdot\bn(\bx) \, \dif\bOmega  \, \dif S(\bx) \right/ (\pi S) 
\end{equation}
where the normalization uses $\int_{\bOmega\cdot\bn(\bx)>0}\bOmega\cdot\bn(\bx)\,\dif\bOmega = \pi$, for any $\bx\in\partial$M, and $S = \int_{\partial\rM}\dif S(\bx)$, which is the surface of the optical medium M.
Indeed, from (\ref{eqn:ct_moms_CF}) with $q = 1$, but for an outgoing flux-based PDF, hence normalization by the uniform boundary irradiation $F_0$ from (\ref{eqn:LT_BCs}), we have
\begin{align} 
\label{eqn:avr_mean_ct}
  \aver{ct} = {}& \left. -\frac{\partial\;}{\partial s} \left. \int_{\partial\rM} \tilde{\bF}(s,\bx)\cdot\bn(\bx) \, \dif S(\bx) \right|_{s=0} \right/ (F_0 S) \nonumber \\
            = {}& \left. \frac{\partial\;}{\partial s} \left. \int_{\rM} \left( \sigma_\ra(\bx) + s \right)\,\tilde{J}(s,\bx) \, \dif V(\bx) \right|_{s=0} \right/ (F_0 S).
\end{align}

The 3D RT problem defined in (\ref{eqn:3D_RTE}) with BCs (\ref{eqn:LT_BCs}) can in fact be solved easily for conservatively scattering media where $\sigma_\ra(\bx) \equiv 0$ (no absorption by particulates, hence $\sigma_\rs(\bx) \equiv \sigma(\bx)$) and in steady state, i.e., $s = 0$ (equivalently, no gaseous absorption, $k_\lambda = 0$).
Indeed, a uniform and isotropic radiance field $\tilde{I}(0,\bx,\bOmega) \equiv I_0 = F_0/\pi$ from (\ref{eqn:LT_BCs}), leading to $\tilde{J}(0,\bx) \equiv J_0 = 4\pi I_0 = 4 F_0$ and $\tilde{\bF}(0,\bx) \equiv \bzero$ verifies (\ref{eqn:3D_RTE}), hence (\ref{eqn:conservation}). 
Moreover, this solution is valid not only for media with arbitrary outer shapes, but also any internal structure as defined by the extinction field $\sigma(\bx)$, including an optical void (where $\sigma(\bx) \equiv 0$).

In this special case, the numerator in (\ref{eqn:avr_mean_ct}) simplifies to $4 F_0 V$, where $V = \int_\rM \dif V(\bx)$ is the volume of M $\subset\Rm^3$.
From there, we recover the known universal result that
\begin{equation} 
\label{eqn:4VoS}
  \aver{ct} = 4V/S
\end{equation}
in 3D, as first proven most elegantly by \cite{BlancoFournier03}.
In 2D RT, all physical units loose a per-m and per-sr becomes per-rad; the isotropic BC in (\ref{eqn:LT_BCs}) becomes $I_0 = F_0/2$ and $J_0 = 2\pi I_0 = \pi F_0$, hence $\aver{ct} = \pi A/P$ where $A$ is the area of M $\subset\Rm^2$ and $P$ is its perimeter.
In \emph{literal} 1D RT, where light can only travel up and down the $z$-axis (as in two-stream theory), physical units all loose another per-m and per-rad goes away; the BC in (\ref{eqn:LT_BCs}) becomes $I_0 = F_0$ and $J_0 = 2 I_0 = 2 F_0$, hence $\aver{ct} = 2 H$ where $H$ is the length of M $\subset\Rm$.
In short, the mean pathlength $\aver{ct}$ is invariably a direct measure of the bulk size of the medium.

\section{Moments of $ct$ for Uniform Optically Thick Plane-Parallel Media}
\label{sec:diff_ppg}

\subsection{Diffusion-Theoretical Model Formulation}

To explore the cloud information content of pathlength moments beyond $\aver{ct}$, starting with $\aver{(ct)^2}$ or $\text{Var}[ct] = \aver{(ct)^2}-\aver{ct}^2$, we consider homogeneous plane-parallel scattering media that are sufficiently opaque for the RT to be treated in the diffusion limit. 
In the diffusion regime, the \emph{exact} (radiant energy) conservation law in (\ref{eqn:conservation}) is complemented with Fick's \emph{approximate} constitutive law:
\begin{equation} 
\label{eqn:constitutive}
  \tilde{\bF} = -\,\frac{1}{3\sigma_\rt}\bnabla\tilde{J},
\end{equation}
where
\begin{equation} 
\label{eqn:scaled_ext}
  \sigma_\rt = (1-g)\sigma_\rs+\sigma_\ra
\end{equation}
is the scaled or ``transport'' extinction coefficient, with $g$ being the scattering PF's asymmetry factor, $2\pi\int_{-1}^{+1}\mu_\rs P(\mu_\rs) \dif\mu_\rs$.
Noting that $\tilde{F}_x = \tilde{F}_y = 0$ in uniformly illuminated plane-parallel, we only need to track $\tilde{F}_z$ (where we will drop the subscript for simplicity).

We have thus reduced the 1D (i.e., homogeneous plane-parallel) RT problem form (\ref{eqn:conservation}) and (\ref{eqn:constitutive}) to the solution of two coupled ordinary differential equations (ODEs),
\begin{align} 
\label{eqn:cons_1D}
   \dif\tilde{F}/\dif z = &{} -(\sigma_\ra+s)\tilde{J}(s,z) \text{ and} \\
\label{eqn:Fick_1D}
   \dif\tilde{J}/\dif z = &{} -3\sigma_\rt   \tilde{F}(s,z), 
\end{align}
subject to BCs
\begin{align} 
\label{eqn:upper_BC}
   \left. (\tilde{J} + 3\chi\tilde{F}) \right|_{z=0} &{} = 4 F_0 \text{ and} \\
\label{eqn:lower_BC}
   \left. (\tilde{J} - 3\chi\tilde{F}) \right|_{z=H} &{} = 0,
\end{align}
where $H$ is the physical thickness of the medium, and $\chi$ is the ``extrapolation'' length factor (typically set to 2/3) that controls the mix of density and current contributions in the above 3$^\text{rd}$-type (a.k.a. Robin) BCs.
We wish to compute the outgoing boundary fluxes
\begin{align} 
\label{eqn:R_dif}
  R(s) = &{} \left. \left. (\tilde{J} - 3\chi\tilde{F}) \right|_{z=0} \right/ (4 F_0)  = \tilde{J}(s,0)/(2 F_0) - 1 \text{ and} \\
\label{eqn:T_tot}
  T(s) = &{} \left. \left. (\tilde{J} + 3\chi\tilde{F}) \right|_{z=H} \right/ (4 F_0)  = \tilde{J}(s,H)/(2 F_0),
\end{align}
respectively for diffuse reflection and total transmission.
We refer to \cite{Davis_etal2009} for derivations.
In the remainder, we assume without loss of generality that $F_0$ is unity.

Those authors also conducted detailed separate investigations of $R(s)$ and $T(s)$ for the conservative scattering case ($\sigma_\ra = 0$) in the limit $s \to 0$ as well as of $\aver{ct}_{R,T}$ and $\aver{(ct)^2}_{R,T}$, from the behavior of $R(s)$ and $T(s)$ as $s \to 0$.
Specifically, considering these quantities as observables (obtained from either active multiple-scattering cloud lidar or passive DO$_2$AS), their dependence on inherent cloud properties was examined from the standpoint of remote sensing.
Key cloud properties targeted are geometric and (scaled) optical thicknesses: $H$ and $\tau_\rt = \sigma_\rt H = (1-g)\sigma H$.
As is well-known, $\tau_\rt$ (hence COT $\tau = \sigma H$, for given $g$) can be derived from $R(0)$ or $T(0)$, respectively from above and below, but sensitivity deteriorates at large $\tau_\rt$.
Specifically, 
\begin{equation} 
\label{eqn:R_cons}
  R(0) = 1-T(0) = \frac{\tau_\rt}{2\chi+\tau_\rt},
\end{equation}
approaches unity, and $T(0) = 1-R(0)$ vanishes, as $\tau_\rt$ increases without bound.
The main new finding is that $\aver{ct}_R$ or $\aver{ct}_T$ can be used to infer $H$, knowing $\tau_\rt$ from (\ref{eqn:R_cons}) or other determinations of COT $\tau$ and $g$.
At large COT (and moderate $\tau_\rt$ if $g = 0.85$, the canonical value for liquid clouds), $R(0)$ looses its sensitivity.
\cite{Davis_etal2009} went on to show, however, that large COT is precisely when $\aver{(ct)^2}_R$ becomes highly sensitive to it, and COT can thus be inferred from this 2$^\text{nd}$-order pathlength moment, knowing $H$ (e.g., from $\aver{ct}_R$) and $g$.
In a sharp contrast, $\aver{(ct)^2}_T$ contains no new cloud information beyond what can be learned from ground-based sensing of $T$ and $\aver{ct}_T$.

\subsection{Mean \emph{and} Variance of $ct$: Non-Absorbing Case}

Here, we are interested in pathlength properties derived from $R(s)+T(s) = 1-A(s)$, where $A(k_\lambda)$ is the bulk gaseous absorption by the medium.
That way, we are effectively assuming that both top and bottom of the slab are irradiated with diffuse light.

\begin{figure}[htbp]
   \centering
   \includegraphics[width=3in]{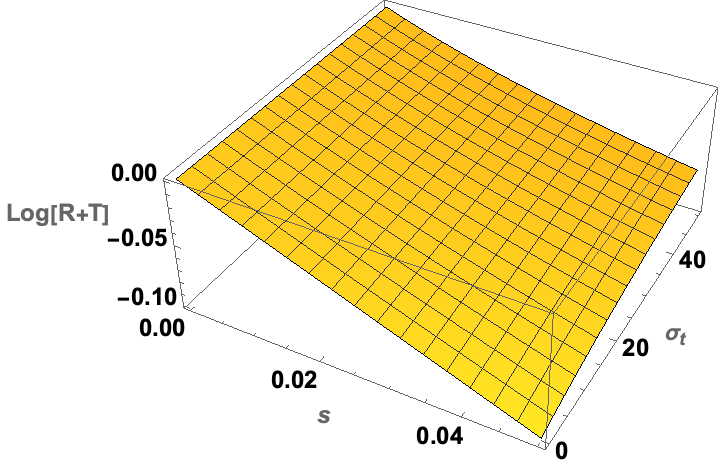}
   \caption{
$\log[R+T]$ from (\ref{eqn:RpT_result}) is plotted as a 2D function of $(s;\sigma_\rt)$ after setting $\chi = 2/3$ and $H = 1$, which can be done without loss of generality. We can, in that case, simply read $\sigma_\rt$ as the scaled COT $\tau_\rt$. 
We note that the slope in $s$ at $s = 0$ is $-$2, independently of the scaled extinction $\sigma_\rt$, that is, the opacity of the medium. 
In (\ref{eqn:RpT_series}) and (\ref{eqn:RpT_moms}), we see that this follows directly from the universal law in (\ref{eqn:4VoS}) for $\aver{ct} = 4V/S$ (= 2$H$ in plane-parallel media).
Curvature of $\log[R+T]$ at $s = 0$ does depend on $\sigma_\rt$, as we will see in (\ref{eqn:RpT_moms})--(\ref{eqn:Var_path}).
   }
   \label{fig:Log_RpT}
\end{figure}

The first step is to solve in closed-form the coupled ODE boundary-value problem in (\ref{eqn:cons_1D})--(\ref{eqn:lower_BC}).
Then, knowing $\tilde{J}(s,z)$, one computes
\begin{equation} 
\label{eqn:RpT_def}
  R(s)+T(s) = \frac{\tilde{J}(s,0)+\tilde{J}(s,H)}{2}-1,
\end{equation}
from (\ref{eqn:R_dif})--(\ref{eqn:T_tot}), leading to:
\begin{equation} 
\label{eqn:RpT_result}
  [R+T](s) = \frac{\sigma_\rt - 3\chi^2 s + 2\chi \sqrt{3 s \sigma_\rt}/\sinh(H \sqrt{3 s \sigma_\rt}) }
                  {\sigma_\rt + 3\chi^2 s + 2\chi \sqrt{3 s \sigma_\rt}/\tanh(H \sqrt{3 s \sigma_\rt}) }.
\end{equation}
Figure~\ref{fig:Log_RpT} shows $\log[R+T](s,\sigma_\rt)$ for $\chi = 2/3.$
We also set $H = 1$ without loss of generality since we can then identify $\sigma_\rt$ with the non-dimensional $\tau_\rt$.
Because of its connection to pathlength moments, we are especially interested in the behavior of $\log[R+T](s,\sigma_\rt)$ for small values of $s$ and arbitrary---but not too small---values of $\sigma_\rt$ ($=\tau_\rt$), to ensure the validity of the diffusion approximation in (\ref{eqn:Fick_1D}).

The next step is to expand $\log[R+T](s)$ in Fig.~\ref{fig:Log_RpT} into a two-term Taylor series at $s = 0$.
Based on (\ref{eqn:RpT_result}), we find
\begin{equation} 
\label{eqn:RpT_series}
  \log[R+T](s) = -3\chi\,H\,s + \frac{3}{4}\,\chi\,(\sigma_\rt H)\,H^2\,s^2 + \mO(s^{5/2}).
\end{equation}
We can identify the terms of that series with those of
\begin{equation} 
\label{eqn:RpT_moms}
  \log[R+T](s) = -\aver{ct}\,s + \frac{1}{2}\,\text{Var}[ct]\,s^2 + \mO(s^{5/2}),
\end{equation}
where we used (\ref{eqn:Avr_ct}) and (\ref{eqn:Var_ct}) in the 1$^\text{st}$- and 2$^\text{nd}$-order terms, respectively.
we note that the 0$^\text{th}$-order term vanishes since $[R+T](0) = 1$.

From the 1$^\text{st}$ terms in (\ref{eqn:RpT_series}) and (\ref{eqn:RpT_moms}), we see that
\begin{equation} 
\label{eqn:Avr_path}
  \aver{ct} = 2H
\end{equation}
for $\chi = 2/3$.
This is as expected from the powerful prediction by \cite{BlancoFournier03} that $\aver{ct} = 4V/S$ for arbitrary \emph{finite-sized} media, as re-derived in the above.
Indeed, starting for a finite parallelepiped of dimensions $W \times W \times H$, its volume is $V = W^2 H$ and its surface is $S = 2W^2+4WH$; so the limit of $4V/S$ as $W\to\infty$ is indeed $2H$, which is also the above prediction for $\aver{ct}$ in literal 1D RT theory using the divergence theorem.

From the 2$^\text{nd}$ terms in (\ref{eqn:RpT_series}) and (\ref{eqn:RpT_moms}), we see that
\begin{equation} 
\label{eqn:Var_path}
  \text{Var}[ct] = \frac{3}{2}\,\chi\,(\sigma_\rt H)\,H^2 = \tau_\rt\,H^2,
\end{equation}
again for $\chi = 2/3$.
It follows from (\ref{eqn:Var_path}) that $\aver{(ct)^2} = \text{Var}[ct] + \aver{ct}^2 = (\tau_\rt+4)\,H^2 \sim \tau_\rt$ when $\tau_\rt \gg 1$. 
That is indeed the predicted scaling by \cite{BlancoFournier06}, $\aver{(ct)^q} \sim \tau_\rt^{q-1} H^q$ (in our notations) for arbitrary optically thick media, in the $q = 2$ case.
In \ref{sec:Appendix_A}, the scaling of $ct$ moments of arbitrary order for reflected light is derived from discrete-time random walk theory, thus reinforcing the robustness of the $ct$-moments/cloud-properties connection.
Moreover, their statement about the prefactor $\partial\aver{(ct)^2}/\partial\tau_\rt = H^2$ agrees with our result for $\chi$ = 2/3.
 
We thus confirm that knowledge of $\aver{ct}$ yields $H$ while knowledge of Var[$ct$] yields $\tau_\rt = (1-g)\tau$, knowing $H$ (from $\aver{ct}$).
We also confirm that the sensitivity of Var[$ct$] to COT $\tau$ is not affected by its value, no matter how large.
The conclusion is the same as obtained by \cite{Davis_etal2009} because, at large COTs where the diffusion approximation works best, $R$ dominates in the sum $[R+T](s)$ at small $s$.
However, the asymptotic ($\tau\to\infty$) partition in the pathlength moments between $R$ and $T$ contributions varies.
Based on individual expressions by \cite{Davis_etal2009}, the $R$-to-$T$ is 2-to-1 for $\aver{ct}$ and 8-to-7 for $\aver{(ct)^2}$.
This rapid trend toward unity makes sense since the higher the moment order, the more it is dominated by the longest paths that typically end in transmission, which compensates for the dwindling weight of the transmitted light, in $2\chi/\tau_\rt$, cf. (\ref{eqn:R_cons}).

\begin{figure}[htbp]
   \centering
   \includegraphics[width=3in]{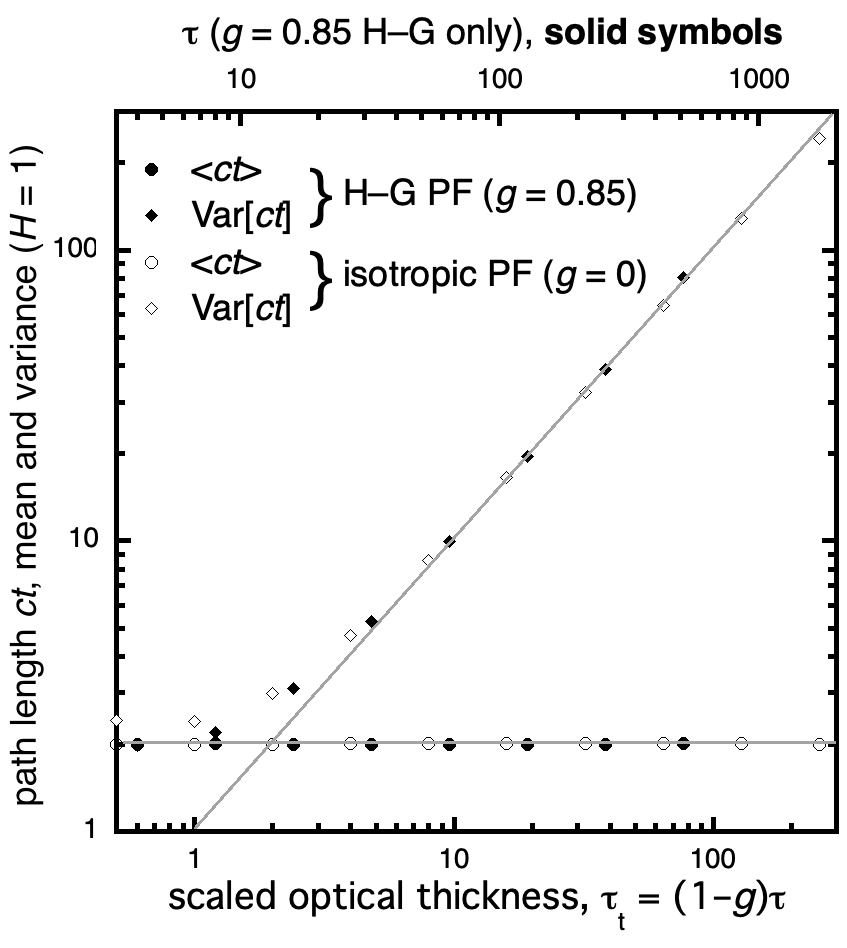}
   \caption{
Numerical validation of predictions for $\aver{ct}$ and Var[$ct$] based on radiative diffusion theory in plane-parallel cloud models. 
The original prediction that $\aver{ct} = 2H$ by \cite{BlancoFournier03} for any COT is verified, as a special case of their general result that $\aver{ct} = 4V/S$ for arbitrary optical media.
Furthermore, our prediction in (\ref{eqn:Var_path}) that Var[$ct$] = $\tau_\rt\,H^2$ is verified in the asymptotic regime ($\tau_\rt \gg 1$).
   }
   \label{fig:Avr_and_Var_ct}
\end{figure}

Numerical validation of the diffusion-theoretical predictions for $\aver{ct}$ and Var[$ct$] in plane-parallel media is in order.
Figure~\ref{fig:Avr_and_Var_ct} shows Monte Carlo simulation results for two choices of scattering PF: isotropic and \cite{H+G1941} (with $g$ = 0.85, as observed in liquid clouds).
COT $\tau$ was varied by factors of 2 from 0.5 to 256 for the isotropic ($g$ = 0) PF, and from 4 to 512 for the Henyey--Greenstein (H--G) PF.
As expected, we see that the $\aver{ct}/H = 2$ prediction is followed exactly to numerical precision.
Also as expected, the Var[$ct$]$/H^2 = \tau_\rt = (1-g)\tau$ prediction is asymptotically exact, and already very good when $\tau_\rt$ exceeds $\sim$4.
This is consistent with the numerical simulations for plane-parallel slabs by \cite{BlancoFournier06} using a linearized Monte Carlo code to compute derivatives with respect to extinction, as displayed in their Table~1.

Equations (\ref{eqn:Avr_path}) and (\ref{eqn:Var_path}), or their more complicated but more relevant counterparts derived by \cite{Davis_etal2009} for reflected light only, can be used to infer cloud properties $(H,\tau_\rt)$ from the mean and variance of $ct$.
However, it is not obvious how to get these pathlength moments from raw DO$_2$AS data.
We therefore describe in \ref{sec:Appendix_B} a tentative algorithm for consideration in existing or future missions.

\subsection{Mean of $ct$: Absorbing Case}

The physical principles of the retrieval of $(H,\tau_\rt)$, a parameterized cloud profile, from reflection DO$_2$AS are very general.
In essence, the light reflected from an optically thick medium is a combination of short and long paths. 
The later dominate the mean pathlength $\aver{ct}$ and the associated light has penetrated the whole cloud, which therefore contains information about the geometrical cloud thickness $H$.
In contrast, pathlength variance Var[$ct$] is determined by the spread between long and short paths.
If the cloud is optically thin, i.e., the \emph{transport} mean-free-path (tMFP) between (effectively isotropic) scatterings, $\ell_\rt = 1/\sigma_\rt$ is commensurate with $H$ and there is not enough space in the medium M to build much diversity in pathlength before the light escapes the medium, so Var[$ct$] is small.
If the cloud is optically thick, $1/\sigma_\rt \ll H$ then the combination of short and long paths is more balanced.
Consequently, Var[$ct$] is large and will increase as $\sigma_\rt$ or $H$ increase since these two parameters respectively make the short paths shorter and the long paths longer.
It is therefore no surprise to find that Var[$ct$] is $\propto \sigma_\rt H = \tau_\rt$.

Cloud particles are non-absorbing at $\Oxy$ A- and B-bands, both in- and out-of-band.
Indeed, only the cloud boundaries are absorbing in the RT problem investigated so far.
However, since the physics behind the inference of $(H,\tau_\rt)$ for an optically thick layer from DO$_2$AS observations is so general, we wonder how much absorption can be tolerated for the retrieval method to still work.
The same question arises for the inference of just $H$ at any optical thickness.
We are motivated here by the potential for DO$_2$AS to be applied to primarily opaque aerosol ``clouds'' of desert dust, wildfire smoke or volcano ash near their sources, where the AOT/COT can be very big.
Can we use DO$_2$AS to go beyond aerosol (layer) top height retrieval, and also obtain its geometrical thickness?

We can begin to address this question theoretically with the same diffusion model used for water clouds as long as we keep absorption small compared to scattering, i.e., $0 < \sigma_\ra \ll (1-g)\sigma_\rs$ in the make up of $\sigma_\rt$ in (\ref{eqn:scaled_ext}).
With that caveat in mind, we can return to the model laid out in (\ref{eqn:cons_1D})--(\ref{eqn:T_tot}).
The quantity of interest from that model, $[R+T](s)$, is obtained in (\ref{eqn:RpT_result}) under the assumption that $\sigma_\ra = 0$ (no particulate absorption). 
By examining (\ref{eqn:cons_1D}), we see that $[R+T](s)$ for the case $\sigma_\ra > 0$ is the same expression, but with $s \mapsto s+\sigma_\ra$.
Therefore, to compute the successive pathlength moments for media with weakly absorbing particles, we only need to expand $\log[R+T](s)$ as it stands in (\ref{eqn:RpT_result}) into a Taylor series at $s = \sigma_\ra$.

\begin{figure}[htbp]
   \centering
   \includegraphics[width=2.6in]{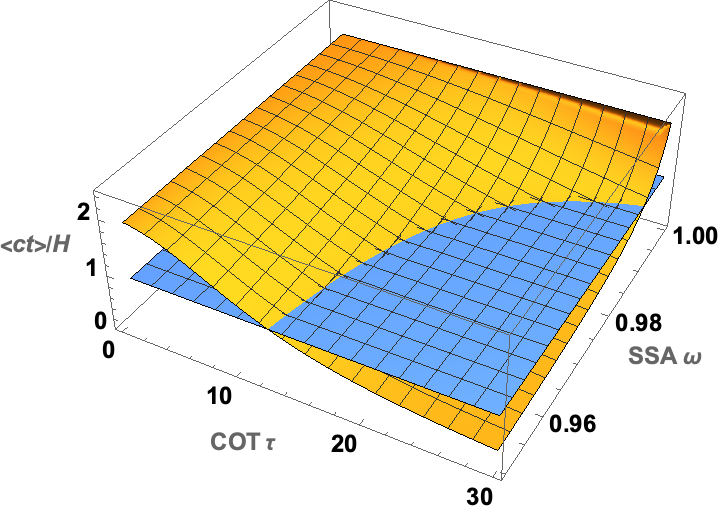}
   \includegraphics[width=2.7in]{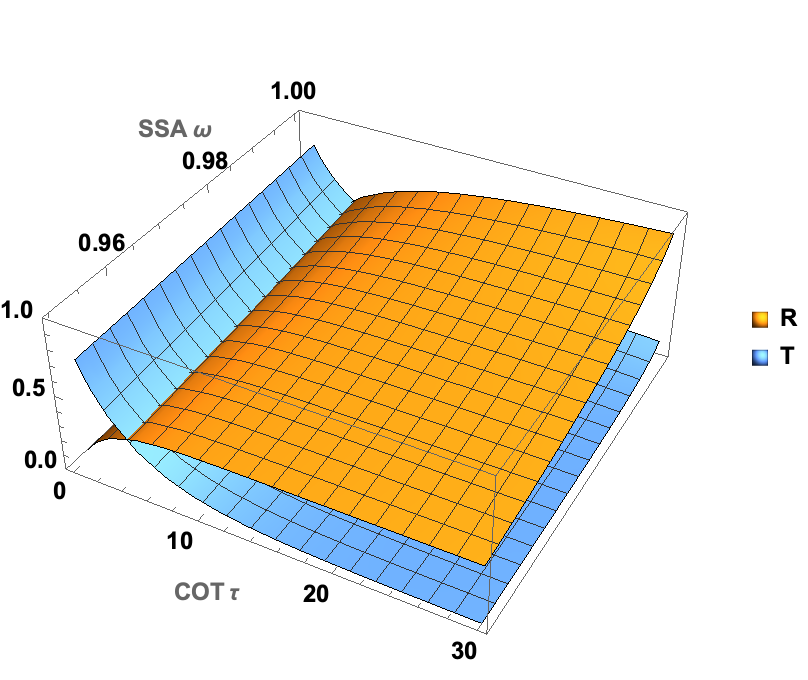}
   \caption{
{\bf Left:} Mean pathlength $\aver{ct}$ in units of layer thickness $H$ for conservative ($\omega = 1$) and absorbing $0.95 \le \omega < 1$ aerosol clouds with varying COT, assuming $g = 0.7$ and $\chi = 2/3$.
The $\aver{ct} = H$ level (1/2 of the $\omega = 1$ case) is plotted for visual reference.
{\bf Right:} Reflectivity $R(\omega,\tau)$ and transmittivity $T(\omega,\tau)$ over the same parameter domain.
   }
   \label{fig:Avr_ct_abs}
\end{figure}

Truncated at 1st-order, the Taylor series of $\log[R+T](s)$ at $\sigma_\ra$ yields (\ref{eqn:RpT_result}) with $s = \sigma_\ra$ at order 0 and, at 1st-order, an expression for $\aver{ct}(\sigma_\ra)$ that is too complicated to be illuminating.
Figure~\ref{fig:Avr_ct_abs} (left panel) shows $\aver{ct}/H$ as a function of the single-scattering albedo (SSA) $\omega = \sigma_\rs/\sigma$ and COT $\tau = \sigma H$ for $g = 0.7$ and $\chi = 2/3$.
The SSA range is limited to [0.95,1] and the aerosol COT varies from 0 to 30, although the diffusion model becomes accurate only past $1/(1-g)$.
A typical value of $g$ for absorbing aerosols is $\approx$0.7, and can be less for fine smoke particles.
The SSA range is limited to [0.95,1] and the aerosol COT varies from 0 to 30, although the diffusion model becomes accurate only somewhat past $1/(1-g) \approx 3.3$ (i.e., $\tau_\rt \approx 1$, about 1/10$^\text{th}$ of the $\tau$-range in Fig.~\ref{fig:Avr_ct_abs}).
The maximum stress for the diffusion model is thus for $\sigma_\ra/\sigma_\rs = 1/\omega-1 \approx 0.05$, which is indeed quite small compared to $1-g = 0.3$.
We see in the left panel of Fig.~\ref{fig:Avr_ct_abs} that $\aver{ct} = 2H$ at $\tau = 0$ for any $\omega$ as well as for $\omega = 1$ for any $\tau$, as shown in the above.
In all other cases, $\aver{ct} < 2H$ since both absorption and opacity reduce pathlengths.
In the left panel of Fig.~\ref{fig:Avr_ct_abs}, we have highlighted the line in $(\tau,\omega)$-space where $\aver{ct} = H$, i.e., half of the pathlength is gone on average.
The lower the SSA, the quicker this happens as COT increases.

An important question arises, from the standpoint of aerosol plume or cloud remote sensing: 
Can we still recover layer thickness $H$ from a DO$_2$AS-derived estimate of in-cloud mean pathlength $\aver{ct}$?
Based on the left panel of Fig.~\ref{fig:Avr_ct_abs}, the answer is affirmative, but the algorithm will need to be informed from other sources about the values of SSA and COT to get the right $\aver{ct}/H$ ratio.

That said, we need to be aware that overall light levels decrease rapidly as the co-SSA $1-\omega$ increases, and that is before adding the gaseous absorption.
The right-hand panel in Fig.~\ref{fig:Avr_ct_abs} quantifies this issue.
Therein, we have have plotted $R(\omega,\tau)$ and $T(\omega,\tau)$ for $g = 0.7$ (most representative of aerosols) and $\chi = 2/3$, from expressions in \cite{Davis_etal2009}.
When $\omega < 1$, $T(\omega,\tau)$ decreases exponentially with $\tau$ at a rate
\begin{equation*} 
  L_\text{dif}(\omega,g) = \frac{1}{\sqrt{3(1-\omega)(1-\omega g)}}, 
\end{equation*}
the so-called diffusion length scale in units where MFP = $1/\sigma$ is unity.
It rapidly goes to $\infty$ as $\omega$ approaches unity, and $T(\omega,\tau)$'s downward trend becomes $\propto$$1/\tau$; for $g = 0.7$ and $\omega = 0.95$, the diffusion scale is only $\approx$4.5 MFPs, $\approx$1/6$^\text{th}$ of the whole $\tau$-range in Fig.~\ref{fig:Avr_ct_abs}.
Meanwhile, $R(\omega,\tau)$ increases at the same exponential rate, asymptoting at
\begin{equation*} 
  R(\omega,\infty) = \frac{1-\chi S(\omega,g)}{1+\chi S(\omega,g)}, \text{ where }S(\omega,g) = \sqrt{\frac{3(1-\omega)}{1-\omega g}}
\end{equation*}
is the so-called similarity factor \citep{King1987}.
For $g = 0.7$ and $\chi = 2/3$, $R(0.95,\infty)$ is only $\approx$0.38, but the reflectivity of a moderately opaque aerosol layer would be even less.
In DO$_2$AS observations, even more light is removed by gaseous absorption but, as shown in Appendix~B, it is fortunately the low values of $k_\lambda$ that are the most informative, at least for the extraction of $\aver{ct}$ and, from there, $H$.



\appendix


\renewcommand{\theequation}{A-\arabic{equation}} 
\setcounter{equation}{0}  

\section{Scaling of Pathlength Moments from Discrete-Time Random Walk Theory at Any Order}
\label{sec:Appendix_A}

It is well-known from the theory of \emph{continuous}-time random walks (RWs), a.k.a. Brownian motion, that the variance $\aver{\bx^2}$ of the RW position increases linearly with time $t$.
Specifically, $\aver{\bx^2}(t) = 6Dt$, resulting from the Gaussian solution of $\partial_t U = D \bnabla^2 U$, subject to $U(0,\bx) = \delta(\bx)$, where $D = c\ell_\rt/3$ is the diffusivity constant in 3D space.
Introducing pathlength $ct$, we have $\aver{\bx^2} = (2\ell_\rt)ct$.
In \emph{discrete}-time RW theory, we investigate the random position $\bx_n$ of the RWing particle after $n \ge 0$ steps.
Since every discrete step corresponds, on average, to a time-step $\ell_\rt/c$, or a pathlength step $\ell_\rt$, we have $ct = \ell_\rt n$, hence
\begin{equation} 
\label{eqn:Var_x_n}
  \aver{\bx_n^2} = (2\ell_\rt^2)n.
\end{equation}
This results directly from the additivity of means and variances over $n$ independent identically-distributed (iid) random variables.
Indeed, the RW is composed of $n$ statistically-uncorrelated (i.e., randomly-oriented) steps, $\bx_i-\bx_{i-1}$, each with variance $2\ell_\rt^2$, starting at $\bx_0 = \bzero$.
Thus, $\bx_n = \sum_{i=1}^n (\bx_i-\bx_{i-1})$, where $\aver{\bx_n} = \aver{\bx_{n-1}} = \dots = \bx_0 = \bzero$, and (\ref{eqn:Var_x_n}) follows for $\aver{\bx_n^2} = \sum_{i=1}^n\aver{(\bx_i-\bx_{i-1})^2}$.

So far, $\bx_n$ is the random variable and $n$ is the given quantity and, so far, the RW unfolds in unbounded 3D space.
Turning the tables, we ask:
On average, what is the number of steps at which a RW starting at $\bx_0 = \bzero$ first reaches the $z = H$ plane?
If the RW is then terminated, it contributes a transmission event in a RW problem over a finite domain 
\begin{equation} 
\label{eqn:slab_H}
  \text{M}(H) = \{\bx\in\Rm^3; 0<z<H\}, 
\end{equation}
so we denote the answer as $\aver{n}_T$.
In the spirit of dimensional analysis, we posit that, based on (\ref{eqn:Var_x_n}), this number surely scales as
\begin{equation} 
\label{eqn:Avr_n_T}
  \aver{n}_T \sim (H/\ell_\rt)^2 = \tau_\rt^2,
\end{equation}
the scale COT from the main text, squared.

How do we now translate this scaling law into information about RWs that end in reflection at $z = 0$?
We need to invoke the lesser-known \emph{law of first returns} \citep{Redner2001}, which assumes the RW is on a half-space, i.e., M($\infty$) in (\ref{eqn:slab_H}), and that the first step is in the positive-$z$ direction.
Let $n$ be the number of steps taken by the RWer before it crosses into the $z < 0$ half-space.
Then Prob$\{n \ge N\} \sim 1/\sqrt{N}$ \citep{Sparre-Anderson1953}, equivalently, 
\begin{equation} 
\label{eqn:Prob_1st_return}
  p_n \sim n^{-3/2},
\end{equation}
since Prob$\{n \ge N\} = \sum_{n=N}^\infty p_n$. 
Note that both mean and variance of $n$, let alone higher moments, are divergent due to the frequent occurrence of RWs that go deep into the positive half-space, even though these RWs will eventually return to $z = 0$ with probability 1.
We also note that (\ref{eqn:Prob_1st_return}) only required that the steps be isotropic \citep[e.g.,][]{FrischFrisch95}; their variance can be infinite, as in so-called L\'evy flights \citep[e.g.,][]{DavisMarshak1997}.

Now, returning to the RWs in the finite slab M($H$), in principle, with $H \gg \ell_t$, we simply truncate the discrete PDF in (\ref{eqn:Prob_1st_return}) for $n > \aver{n}_T$ from (\ref{eqn:Avr_n_T}).
In that case,
\begin{equation} 
\label{eqn:n_moms_R_def}
  \aver{n^q}_R = \left. \sum_{n=1}^{[\aver{n}_T]} n^q p_n \right/ \sum_{n=1}^{[\aver{n}_T]} p_n,
\end{equation}
where $q \ge 0$ and $[\cdot]$ designates the closest integer value.
For $\aver{n}_T$, that is, $\tau_\rt^2$ from (\ref{eqn:Avr_n_T}), sufficiently large, the denominator in (\ref{eqn:n_moms_R_def}) is close to unity since we assume that the PDF in (\ref{eqn:Prob_1st_return}) is normalized.
Consequently,
\begin{equation} 
\label{eqn:n_moms_R_result}
  \aver{n^q}_R \approx \sum_{n=1}^{[\tau_\rt^2]} n^{q-3/2} \sim (\tau_\rt^2)^{q-1/2} = \tau_\rt^{2q-1}.
\end{equation}
From there, we can derive the anticipated scaling for
\begin{equation} 
\label{eqn:ct_moms_R}
  \aver{(ct)^q}_R \sim \tau_\rt^{2q-1}\,\ell_rt^q = H^q \times  \tau_\rt^{q-1},
\end{equation}
using $\ell_rt = H/\tau_\rt$
Equation~(\ref{eqn:ct_moms_R}) is exactly as predicted by \cite{BlancoFournier06}, but in our notations.
In particular, we retrieve $\aver{ct}_R \sim H$ and $\aver{(ct)^2}_R \sim H^2\,\tau_\rt$.

\renewcommand{\theequation}{B-\arabic{equation}} 
\setcounter{equation}{0}  

\section{Algorithm for Deriving Low-Order Pathlength Moments from DO$_2$AS Data}
\label{sec:Appendix_B}

In the presence of multiple scattering, spectroscopic observations with real-world instruments involve not one but two averages at a miniumum.
Even in the line-by-line limit (i.e., ultra-fine spectral resolution), one must average over all possible pathlengths $L = ct$, as in (\ref{eqn:LT_I}), from the main text, where $s = k_\lambda$.
The underlaying pathlength distribution in (\ref{eqn:LT_I}) is precisely the target of the study since it contains information about the cloud.

Real spectrometers have finite spectral resolution $\Delta\lambda$, and inside a spectral bin $[\lambda-\Delta\lambda/2,\lambda+\Delta\lambda/2)$ one can potentially find many values of the molecular absorption coefficient $k_\lambda$.
The instrument is therefore performing a de facto integration over all wavelengths $\lambda$ in that interval, which is equivalent to an integration over a whole distribution of $k_\lambda$ values.
In RT simulations, this last spectral integration is commonly approximated by a weighted sum, $\int_{\lambda-\Delta\lambda/2}^{\lambda+\Delta\lambda/2}\Phi(k_\lambda)\dif\lambda \approx \sum_n g_n(\lambda,\Delta\lambda) \Phi(k_\lambda)$, where the number of terms as well as the weights will depend on the diversity of $k_\lambda$ values in $[\lambda-\Delta\lambda/2,\lambda+\Delta\lambda/2)$.

Suppose now that we are given DO$_2$AS ratio data in the form of
\begin{equation} 
\label{eqn:DOAS_ratio}
  r_\text{DO2AS}(\lambda,\Delta\lambda;\cdots) = \frac{1}{\tilde{I}(0,\cdots)} \sum\limits_n g_n(\lambda,\Delta\lambda) \tilde{I}(k_\lambda,\cdots), 
\end{equation}
for $\lambda_\text{min} < \lambda < \lambda_\text{max}$, hopefully with several if not many spectral samples.
It will likely be based on some averaging and/or sampling of the transport space coordinates $(\bx,\bOmega)$, e.g., $\bx$ spans an imager pixel and $\bOmega$ is some viewing direction; we have represented this averaging/sampling here with ``$\cdots$'' as a placeholder.
The spectral signal in (\ref{eqn:DOAS_ratio}) fluctuates up and down, starting at the continuum where $r_\text{DO2AS}(\lambda,\Delta\lambda;\cdots)$ is unity because $k_\lambda$ vanishes there by definition.
The first order-of-business is to produce a scatter plot the $r_\text{DO2AS}(\lambda)$ data versus the known values of $k_\lambda$. 
This scatter plot starts a unity (with $k_\lambda$ = 0) and trends downward to a minimum associated with the largest value of $k_\lambda$ in the spectral window $[\lambda_\text{min},\lambda_\text{max}]$.

As previously stated, the individual spectral channels will in general be wide enough to contain a possibly large range of $k_\lambda$ values in the line-by-line sense.
It is therefore key to assign a meaningful \emph{effective} value $k_\lambda^\text{eff}$ to each spectral channel.
There is no unique way of doing this, but a rational approach is to use the same weights as in (\ref{eqn:DOAS_ratio}): $k_\lambda^\text{eff}(\Delta\lambda) = \sum_n g_n(\lambda,\Delta\lambda) k_\lambda$.
Based on (\ref{eqn:ct_moms_CF}) in the main text, the forward model for the DO$_2$AS ratio data in the form of a truncated Taylor series expansion at $k_\lambda = 0$ is then
\begin{equation} 
\label{eqn:DOAS_ratio_fit}
  r_\text{DO2AS}(\lambda,\Delta\lambda;\cdots) = 1 - \aver{L}\,k_\lambda^\text{eff}(\Delta\lambda) + \frac{\aver{L^2}}{2}\,\left( k_\lambda^\text{eff}(\Delta\lambda) \right)^2 + \dots
\end{equation}

Based on (\ref{eqn:DOAS_ratio_fit}), the algorithm to extract $\aver{L}$ and $\aver{L^2}$ from DO$_2$AS observations is as follows.
\begin{itemize}
\item Generate scatter plot of $\left( k_\lambda^\text{eff}(\Delta\lambda), r_\text{DO2AS}(\lambda,\Delta\lambda;\cdots) \right)$ pairs where at least one data point is out-of-band (i.e., $k_\lambda^\text{eff}(\Delta\lambda)$ = 0 and $r_\text{DO2AS}(\lambda,\Delta\lambda;\cdots)$ = 1).
\item Apply to these data a constrained least-squares polynomial fit of order at least order 2, the constraint being to go through the (0,1) point.
\item Identify the 1st-order coefficient with $\aver{L}$.
\item Identify the 2nd-order coefficient with $\aver{L^2}$/2.
\end{itemize}
Depending on the total number of available data points, hopefully more than 3, one can fit a polynomial of order $>$2 if it helps to reduce the uncertainty on the first two coefficients.
To obtain cloud properties from these two pathlength moments, we can use (\ref{eqn:Avr_path}) and (\ref{eqn:Var_path}) from the main text.
However, their simplicity comes from the simplifying assumption of uniform (top \emph{and} bottom) illumination, hence a mix of reflected and transmitted light.
\cite{Davis_etal2009} offer more involved formulas for the two pathlength moments that are more realistic since derived only for the observed (namely, reflected) light and, moreover, account for important parameters such as the solar zenith angle.

 \bibliographystyle{elsarticle-num} 
 \bibliography{references}






\vspace{12pts}
\copyright 2021. All rights reserved.

\end{document}